\documentclass[showpacs,a4paper,preprint,titlepage,secnumarabic,groupedaddress,prl]{revtex4}
\usepackage{amssymb}
\usepackage{amsmath}
\usepackage[final]{graphicx}
\usepackage{dcolumn}
\usepackage{bm}
\usepackage{verbatim}

\input{tcilatex}
\begin{document}

\title{Complex oscillatory yielding of model hard sphere glasses}
\author{N. Koumakis$^{1}$, J. F. Brady$^{2}$ and G. Petekidis$^{1,\ast }$}
\affiliation{$^{1}$FORTH/IESL and Department of \ Materials Science \& Technology,
University of \ Crete, 71110, Heraklion, Greece}
\affiliation{$^{2}$Division of Chemistry and Chemical Engineering, Caltech, Pasadena, CA
91125, USA}

\begin{abstract}
The yielding behaviour of hard sphere glasses under large amplitude
oscillatory shear has been studied by probing the interplay of Brownian
motion and shear-induced diffusion at varying oscillation frequencies.
Stress, structure and dynamics are followed by experimental rheology and
Browian Dynamics simulations. Brownian motion assisted cage escape dominates
at low frequencies while escape through shear-induced collisions at high
ones, both related with a yielding peak in\ $G^{\prime \prime }$. At
intermediate frequencies a novel, for HS glasses, double peak in $G^{\prime
\prime }$ is revealed reflecting both mechanisms. At high frequencies and
strain amplitudes a persistent structural anisotropy causes a stress drop
within the cycle after strain reversal, while higher stress harmonics are
minimized at certain strain amplitudes indicating an apparent harmonic
response.
\end{abstract}

\date{29/1/13}
\maketitle

Hard Sphere (HS)\ colloids have been used as model systems to study\ a
plethora of fundamental condensed matter problems such as the interplay
between equilibrium phases (crystals and liquids) and non-ergodic states
(glasses or gels), and their behavior under external fields such as shear%
\cite{MewisWagnerBook,LarsonBook}. A major goal is to develop an
understanding that will enable tailoring of the mechanical and flow
properties based on the structure and dynamics at the particle level. HSs
form metastable glasses above a volume fraction of about $\varphi _{g}\sim
0.59$ where crystallization and long-time diffusion are suppressed\cite%
{HSNature,Puseyrev}. Such states exhibit solid-like behavior at low
stresses/strains and shear-melting (yielding) above a yield stress/strain%
\cite{Petekidis04}. This phenomenon is investigated more effectively with a
combination of rheological and optical or scattering techniques that may
unravel the\ link between microstructure, dynamics and mechanical properties
in steady\cite{besseling07,Koumakis12b} or oscillatory shear \cite%
{Petekidis02,Petekidis03}. Such studies have shown that particle cages are
deformed under shear and particles move irreversibly when a critical strain
is exceeded, leading to flow.

Experimentally, large amplitude oscillatory shear tests are widely used to
monitor yielding due to their simplicity and relation with linear elastic
and viscous moduli, $G^{\prime }$ and $G^{\prime \prime }$. However, their
interpretation in the non-linear regime becomes complex due to stress
distortion introducing higher harmonics\cite{HyunReview}. In a wide range of
systems (emulsions, polymers and colloids) a generic peak in $G_{1}^{\prime
\prime }$\ (viscous modulus of the fundamental frequency) is observed
representing an increased energy dissipation near the yield strain, $\gamma
_{y}$, where $G_{1}^{\prime }=G_{1}^{\prime \prime }$, beyond which the
sample flows\cite{Mason95,Petekidis03}. On the theory side, Mode Coupling
Theory (MCT)\cite{BraderFuchsPRE10} as well as the semi-phenomenological
Soft Glassy Rheology (SGR) \cite{SollichSGR} are able to capture some
aspects of Large Amplitude Oscillatory Shear (LAOS) such as the peak of $%
G_{1}^{\prime \prime }$. Nevertheless, the underlying mechanisms relating
stress with shear induced structure such as cage deformation, breaking and
reformation, as well as particle displacements, are poorly understood, and
the frequency dependence is largely unexplored.

LAOS experiments are expected to provide valuable information on the
interplay of Brownian motion and shear during yielding of HS glasses. Along
this line light scattering-echo experiments probing the average particle
displacements under oscillatory shear\cite{Petekidis02,Petekidis03} revealed
a transition to irreversibility beyond a critical strain amplitude\cite%
{Petekidis02,Petekidis03}. This can be viewed as the analogue of the shear
induced irreversibility observed in concentrated non-Brownian particles\cite%
{Pine2005}, since in colloidal glasses out-of-cage diffusion is frozen,
while in non-Brownian particles diffusion is absent at all length scales.
While experiments combining rheometry with scattering or microscopy are
rather demanding \cite%
{Petekidis02,Ballesta2008,Besseling2010,LopezWagnerPRL12,LettingaPRL12},
computer simulations can provide an alternative route.

Here we use a combination of oscillatory shear rheometry and Brownian
Dynamics (BD) simulations to investigate the links between structure and
particle dynamics with the non-linear rheological response of HS glasses in
a wide range of frequencies, $\omega $, non-dimensionalized by $Pe_{\omega
}^{0}$ $=\omega \tau _{B}$, with $\tau _{B}=R^{2}/D_{0}$, $R$ the radius and 
$D_{0\text{ }}$the free diffusion coefficient, or $Pe_{\omega }$ if the
short-time self diffusion coefficient $D_{s}(\varphi )$ is$\ $used. At low $%
Pe_{\omega }^{0}$ yielding is related to Brownian-assisted irreversible
particle motion manifested in a Dynamic Strain Sweep (DSS) with the peak of $%
G_{1}^{\prime \prime }$. However, at the largely unexplored regime of high $%
Pe_{\omega }^{0}$, we detect collision-dominated yielding and, within the
oscillation period, a strongly anisotropic structure causing a reduced
stress beyond strain reversal (memory of structure due to lack of Brownian
relaxation). At intermediate $Pe_{\omega }^{0}$, the sample is affected by
both mechanisms as manifested by a novel, for HS glasses, double peak in $%
G_{1}^{\prime \prime }$.

We used sterically stabilized poly(methyl methacrylate) (PMMA) model nearly
hard-sphere particles with two radii, $R=358$ $\,\mathrm{nm}$ and $130\ $%
\textrm{nm} (polydispersity $10-12\%$) suspended in octadecene and
octadecene/bromonapthalene mixture respectively, in order to expand the $%
Pe_{\omega }$ range. We prepared different volume fractions, $\varphi $, by
diluting a random close packed batch with the exact $\varphi $ then adjusted
by matching $G_{1}^{\prime }$ of the two systems in agreement with \cite%
{Koumakis12a}. Experiments were performed on an ARES strain controlled
rheometer (with 25mm diameter/0.01rad angle cone-plate) and a solvent trap
to eliminate evaporation. In BD simulations, HS interactions were
implemented through the potential free algorithm\cite{FossBrady}.
Oscillatory shear was applied with periodic boundary conditions using
typically $5405$ particles with $10\%$ polydispersity to avoid
crystallization.

In figure 1 we show DSS tests performed at low and high $Pe_{\omega }$,
together with the evolution of the normalized intensity of all higher stress
harmonics, $I_{all}/I_{1}=\sum I_{i}/I_{1}$, ($i=2n+1$, $n\geq 1$) , as well
as Lissajous curves (intracycle stress versus strain) for representative
strain amplitudes, $\gamma _{0}$. The low$\ Pe_{\omega }\ (=0.5$) data (fig.
1a) exhibit the typical DSS response with a $G_{1}^{\prime \prime }$ peak
around $\gamma _{y}$ (at $G_{1}^{\prime }=G_{1}^{\prime \prime }$) similar
to previous studies\cite{Mason95,Pham08,Koumakis12a}. The non-linear
response is accompanied by progressively larger intracycle non-linearities
as indicated both by the Lissajous curves and the increasing $I_{all}/I_{1}$
as expected\cite{HyunReview,Koumakis12a}. The latter increases beyond
yielding and reaches almost $30\%$ at high $\gamma _{0}$ as found previously%
\cite{Koumakis12a}. Moreover, the Lissajous plots show a transition from a
linear viscoelastic behavior (elliptical shape) at low $\gamma _{0}$, to a
parallelogram pattern indicative of a intracycle sequence of elastic-plastic
response at $\gamma _{0}>\gamma _{y}$ \cite{Koumakis12a,HyunReview}.

\begin{figure}[ptb]\begin{center}
\includegraphics[
natheight=3.3814in, natwidth=5.0125in, height=3.4091in, width=5.0401in]
{c:/Users/Georgp/Documents/PapersText/Papers/Submitted/LAOS_Nick_short/final/graphics/figure1_v6__1.pdf}%
\caption{Top: Dynamic strain sweeps for (a) $R=130$ nm, $\protect\varphi %
=0.60$ at $\protect\omega =1$rad/s ($Pe_{\protect\omega }^{0}=0.04$, $Pe_{%
\protect\omega }=0.5$ ) and (b) $R=358$ nm $\protect\varphi =0.60$ at $%
\protect\omega =1$ rad/s ($Pe_{\protect\omega }^{0}=0.9$, $Pe_{\protect%
\omega }=11.2$), with the 1st harmonic of the elastic $G_{1}^{\prime }$ and
viscous $G_{1}^{\prime \prime }$\ modulus as a function of strain amplitude.
Middle: Representative Lissajous plots are shown in different strain
amplitudes as indicated. Bottom: Normalized total intensity of the higher
harmonics of the stress, $I_{all}/I_{1}$.}\label{figure 1}%
\end{center}\end{figure}%

At high $Pe_{\omega }$, achieved with large particles ($R=358$ nm) at $%
\omega =1$ rad/s ($Pe_{\omega }=11.2$) the response is qualitatively
different (fig. 1b). Firstly, the peak of $G_{1}^{\prime \prime }$ shifts to
higher $\gamma _{0}$, beyond $\gamma _{y}$. Secondly, $I_{all}/I_{1}$ (and
individual $I_{2n+1}/I_{1}$) exhibits a non-monotonic behavior showing a
first maximum around $\gamma _{y}$ and subsequently decreases substantially
well inside the non-linear regime. Hence, the sample exhibits a more
harmonic stress response (anharmonicity is lowered) even though under
non-linear LAOS.

We further explore the $Pe_{\omega }$ dependence by changing $\omega $ while
keeping $\gamma _{0}$ constant. In fig. 2a experiments with varying $\omega $
(at $\gamma _{0}=100\%)$ reveal the transition from the low $Pe_{\omega }$
rectangular shaped Lissajous curves reflecting a sequence of elastic and
plastic responses, to the high $Pe_{\omega }$ regime with a characteristic 
\textit{ellipsoid with a double concave distortion} caused by reduced stress
in the quandrants II and IV after strain reversal. The intensity\ of the 3rd
harmonic at $100\%$ (above $\gamma _{y}$) exhibits a minimum with $\omega $
in experiments, $\varphi =0.62$, and BD, $\varphi =0.60$, (fig. 2c); note
that the position of minimum is $\varphi $ and $\gamma _{0}$ dependent.
While in both $Pe_{\omega }$ regimes the stress response is highly
anharmonic, with significant $I_{3}/I_{1}$, during the transition the
Lissajous curves (fig. 2a) acquire an ellipsoid shape involving almost zero
higher harmonic contributions. BD simulations showing identical rheological
response (fig. 2b) with experiments are able to provide valuable structural
information revealing the underlying mechanism of such stress reduction. In
fig. 2d we plot the 2D projection of the pair correlation function in the
velocity-gradient (xy) direction, $g_{xy}(r)$, at specific points inside the
oscillation cycle for $Pe_{\omega }=100$, similar to findings under steady
shear\cite{Koumakis12b}. Contrary to what we find at low $Pe_{\omega }$
(supplemental material), here the structure is highly anisotropic at the
point of maximum strain (zero shear rate; point A in fig. 2b). Moreover,
such anisotropy is persistent during a large part of the successive
quandrant where the shear has been reversed (points B and C). Such an
anisotropic cage, created during high $Pe_{\omega }$ shear in one direction,
allows flow with less stress (due to fewer particle collisions) when shear
is reversed. Thus, in quandrants II and IV the stress is reduced if compared
to a fully harmonic viscous response corresponding to the flowing
anisotropic structure of quadrants I and III (dash-dot line in fig. 2b).
Only beyond zero strain (points D and E) is the structure reversed and the
stress comes back to the maximum values within the period (see supplemental
material). Such response is absent at low $Pe_{\omega }$ since Brownian
motion relaxes shear-induced structural anisotropy more efficiently, and the
stress response is similar in all quandrants.

\begin{figure}[ptb]\begin{center}
\includegraphics[
natheight=8.2538in, natwidth=11.6949in, height=4.5463in, width=6.4307in]
{c:/Users/Georgp/Documents/PapersText/Papers/Submitted/LAOS_Nick_short/final/graphics/figure2_v8__2.pdf}%
\caption{Lissajous plots at $100\%$ strain amplitude (a) for experiments at
different $Pe_{\protect\omega }$ denoted by the dash lines in (c) with the 4
quantrants of the oscillatory cycle indicated and (b) BD simulations at $Pe_{%
\protect\omega }=100$. The dashed line in (b) indicates a viscous harmonic
stress strain response (see also supplemental material). (c) $Pe_{\protect%
\omega }$ dependence of the 3rd harmonic at $100\%$, for experiments at $%
\protect\varphi =0.62$ ($R=358$ nm) and BD simulations at $\protect\varphi %
=0.60$. (d) 2D projections in the velocity-gradient (xy) plane of the
difference of $g_{xy}(r)$ under shear from that at rest from BD\ simulations
at indicated points within the cycle, points A-E in (b).}\label{figure 2}%
\end{center}\end{figure}%

The data presented above verify the existence of the two $Pe_{\omega }$
regimes: The low $Pe_{\omega }$ one, conventionally studied up to now, where
Brownian motion is dominant and the high $Pe_{\omega }$ where shear-induced
particle collisions introduce novel LAOS features related to the persistent
structural anisotropy and the consequent reduced stress after strain
reversal. The transition from Brownian activated yielding, where particles
under shear escape their cages assisted by thermal motion, to
collision-induced cage breaking at high frequencies is linked to pronounced
irreversible particle rearrangements and decreasing $\gamma _{y}$ at low
frequencies\cite{Petekidis02,Petekidis03}.

\begin{figure}[ptb]\begin{center}
\includegraphics[
natheight=5.6109in, natwidth=4.9978in, height=5.6386in, width=5.0254in]
{c:/Users/Georgp/Documents/PapersText/Papers/Submitted/LAOS_Nick_short/final/graphics/figure3_v6__3.pdf}%
\caption{(a) Dynamic strain sweeps with $R=130$ nm particles at $\protect%
\varphi =0.639$ and an intermediate ($\protect\omega =10$ rad/s, $Pe_{%
\protect\omega }^{0}=0.4$, $Pe_{\protect\omega }=8$ thick black lines) and
low ($\protect\omega =1$ rad/s, $Pe_{\protect\omega }^{0}=0.04$, $Pe_{%
\protect\omega }=0.8$ thin red lines) frequency regime.\ (b) Indicative
Lissajous plots for $Pe_{\protect\omega }=8$ at different strain amplitudes
and (c) Normalized total intensity of higher harmonics $I_{all}/I_{1}$%
,versus strain amplitude for $Pe_{\protect\omega }=8$.}\label{figure 3}%
\end{center}\end{figure}%

We further investigated HS glasses at higher $\varphi $ and frequencies
corresponding to an intermediate ($\varphi $ dependent) $Pe_{\omega }$
regime. Experimental DSS, Lissajous plots and $I_{all}/I_{1}$ shown in fig.
3 as a function of $\gamma _{0}$ for $\varphi =0.639$ reveal an even richer
mechanical response at $Pe_{\omega }^{0}=0.4$ $(Pe_{\omega }=8)$. A main
observation here is the unambiguous detection of two peaks in $G_{1}^{\prime
\prime }$\ detected for the first time in HS glasses. Such a feature was so
far observed only in attractive glasses and gels indicating a two-step
yielding due to two length-scales present in attractive systems i.e. the
interparticle bond and the cage or cluster size\cite{Pham08,Koumakis11}.
However, the double $G_{1}^{\prime \prime }$ peak seen here must be of a
different nature since a second length-scale is absent and moreover, the
phenomenon is only observed in a narrow range of $Pe_{\omega }$. The first
peak of $G_{1}^{\prime \prime }$ is identified with the one observed at low $%
Pe_{\omega }$ by a direct comparison of the two DSSs ($Pe_{\omega }^{0}=0.04$
and $0.4)$ shown in fig 3a. The two peaks signify the maximum in energy
dissipation during the two yielding mechanisms at low and high $Pe_{\omega
}^{0}$, attributed to cage breaking via shear-assisted activated hopping and
through particle collisions, respectively. The Lissajous figures and higher
harmonics reveal a transition from the low $\gamma _{0}$ linear response to
a viscoplastic flow at high $\gamma _{0}$ passing through two states with
apparent harmonic response as indicated by the two minima of $I_{all}/I_{1}$%
. At these strain amplitudes the Lissajous curves acquire an nearly
ellipsoidal shape due to the compensation of the structural phenomena during
the transition from low to high $Pe_{\omega }$ (as in fig 2c). Note that for
this sample ($R=130$ nm, $\varphi =0.639$) the high $Pe_{\omega }$ (fig 1b)
was not within the experimental window.

\begin{figure}[ptb]\begin{center}
\includegraphics[
natheight=3.3961in, natwidth=5.0004in, height=3.4238in, width=5.028in]
{c:/Users/Georgp/Documents/PapersText/Papers/Submitted/LAOS_Nick_short/final/graphics/figure4_v6__4.pdf}%
\caption{BD\ simulations at $\protect\varphi =0.60$: (a) DSS data showing $%
G_{1}^{\prime }$ and $G_{1}^{\prime \prime }$ as a function of $\protect%
\gamma _{0}$ for different $Pe_{\protect\omega }$ as indicated. (b) Power
law exponent for the $\protect\gamma _{0}$ dependence of $G_{1}^{\prime
\prime }$ and $D_{eff}(t=T)$ as a function of $Pe_{\protect\omega }$ (c)
Average $D_{eff}(t=T)$ versus $\protect\gamma _{0}$ for different $Pe_{%
\protect\omega }$ and (d) $D_{eff}(t=T)$ versus $Pe_{\protect\omega }$ for $%
\protect\gamma _{0}=10\%$, $30\%$ and $100\%$. The corresponding $%
D_{eff}(t=T)$ for a system at rest is also indicated.}\label{figure 4}%
\end{center}\end{figure}%

Fig. 4a shows $G_{1}^{\prime }$ and $G_{1}^{\prime \prime }$ from BD LAOS
tests for $\varphi =0.60$ at different $Pe_{\omega }$. The dependence of \ $%
G_{1}^{\prime }$ and $G_{1}^{\prime \prime }$ at high $\gamma _{0}$ ($%
>\gamma _{y}$) follows a power law decrease $G_{1}^{\prime }$($G_{1}^{\prime
\prime }$)$\varpropto $ $\gamma _{0}^{\nu ^{\prime }}$ ($\gamma
_{0}^{v^{\prime \prime }}$) as detected experimentally\cite%
{Mason95,Koumakis12a}. While Maxwell-type models give $\nu ^{\prime }=2\nu
^{\prime \prime }=-2$ and MCT (around the glass transition) predicts lower
values but similar $\nu ^{\prime }/\nu ^{\prime \prime }$ ratio \cite%
{Miyazaki06,BraderFuchsPRE10} experiments in HS glasses show deviations from
such simple dependency \cite{Koumakis12a}. In agreement with experiments
(fig. 1b and supplemental material), BD simulations give $Pe_{\omega }$
dependent exponents (fig 4a) with $\nu ^{\prime \prime }$ approaching $-1$ \
and $0$ at low and high $Pe_{\omega }$ respectively. Therefore, at high $%
Pe_{\omega }$, where collision activated out-of-cage particle rearrangements
are dominant, $\gamma _{0}G_{1}^{\prime \prime }\ \simeq \sigma (\overset{%
\cdot }{\gamma }_{\max })$, a measure of energy dissipation per unit strain,
is proportional to shear rate ($\gamma _{0}\omega $) similar to the limiting
high shear rate viscosity behavior under steady shear. On the other hand at
low $Pe_{\omega }$ a\ $G_{1}^{\prime \prime }\varpropto $ $\gamma _{0}^{-1}$
dependence corresponds to a steady shear shear thinning response of a HS
glass at the yield stress plateau with $\eta _{eff}$ $\propto \overset{\cdot 
}{\gamma }^{-1}$. Note that at this stage, the absence of a double $%
G_{1}^{\prime \prime }$ peak in BD can not be firmly attributed to the
absence of hydrodynamic interactions, since BD at high volume fractions
could not be conducted at a similar number of data points as experiments due
to computational time restrictions.

Further insight in the two yielding mechanisms is gained by examining
microscopic particle dynamics within a LAOS cycle by BD. In Fig. 4c we show
the effective diffusivity, $D_{eff}(t=T)=\langle \Delta z^{2}(T)\rangle /T$
with $\langle \Delta z^{2}(T)\rangle $ the mean square displacement in the
vorticity direction, $z$, (at $t=T$ ), as a function of $\gamma _{0}$ for
several $Pe_{\omega }$. At low $Pe_{\omega }$, $D_{eff}(T)$ increases
sublinear with $\gamma _{0}$, whereas as $Pe_{\omega }$ is increased it
exhibits progressively a weaker increase at small $\gamma _{0}$ and a
stronger one at higher. The constant $D_{eff}(T)$ at low $\gamma _{0}$
indicates that prior to yielding in-cage diffusion is unaffected by shear.
For $\gamma _{0}>\gamma _{y}$, $D_{eff}(T)$, corresponding to out-of cage
diffusion, increases sublinear ($\gamma _{0}=100\%$, fig. 4d) with a power
law exponent $\simeq 0.8$ at low $Pe_{\omega }$, which approaches $1$ for $%
Pe_{\omega }>1$ where direct particle collisions are dominant. In
comparison, the corresponding $D_{eff}(T)\ $for\ $\gamma _{0}=10\%$ and $%
30\% $ show an initial sublinear dependence at low $Pe_{\omega }$ and
subsequently approach the curve at rest, since for such frequencies $\gamma
_{y}>30\%$ \cite{Petekidis03} and the sample has not yet yielded. Note that
a similar power law increase (with exponent $\sim 0.8)$ has been detected in
HS\ glasses at low steady shear rates by confocal microscopy \cite%
{besseling07}, a weak but systematic deviation from the linear MCT prediction%
\cite{FuchsCatesJoR} and closer to agreement with\ non-linear Langevin
equation theory involving activated hoping mechanisms \cite{SaltzmanJPCM}.
For our oscillatory BD a linear dependence is reached\ for $Pe_{\omega }>10$
as expected for non-Brownian particles under steady shear\cite{SierouBrady04}%
. We suggest that the sublinear and linear dependencies probed here at low
and high $Pe_{\omega }$, respectively, reflect the two different mechanisms
involved in the two regimes. Note that similarly with $D_{eff}$, $\gamma
_{0}G_{1}^{\prime \prime }$ increases linearly with $\gamma _{0}$ at high $%
Pe_{\omega }$; the linear $\gamma _{0}$ dependence of both quantities is
linked to collision-induced yielding. On the other hand at low $Pe_{\omega }$
both quantities increase sublinearly, $D_{eff}\varpropto $ $\gamma
_{0}^{0.8} $ and $\gamma _{0}G_{1}^{\prime \prime }\varpropto $ $\gamma
_{0}^{0.4}$ (the latter exponent tends to zero as $Pe_{\omega }$ is lowered)
reflecting plastic flow, with a power law stress behavior, and complex
Brownian/shear-activated particle hoping.

In summary, the combination of experimental oscillatory rheology and BD
simulations has revealed the complete mechanical fingerprint of HS glasses
and the related underlying microscopic structure and dynamics over a wide
frequency regime where both Brownian and non-Brownian behavior is probed. At
low $Pe_{\omega }$, commonly studied up to now, Brownian-assisted
irreversible motion takes place during yielding with a single peak of $%
G_{1}^{\prime \prime }$ and strong higher harmonics of the stress at large $%
\gamma _{0}$. In this regime the microstructure under shear is only weakly
anisotropic while the shear-induced diffusivity scales sublinear with $%
Pe_{\omega }$. At high $Pe_{\omega }$ yielding is dictated by
collision-induced displacements linked with a shear-induced long time
diffusion that increases linearly with $Pe_{\omega }$. A single $%
G_{1}^{\prime \prime }$ peak is detected at $\gamma _{0}>\gamma _{y}$ which
eventually turns into a plateau at the limit of large $Pe_{\omega }$. The
structure under shear is strongly anisotropic but more interestingly
exhibits a hysteresis under strain reversal. Such structural memory due to
lack of Brownian relaxation, causes a characteristic stress drop after
strain reversal as revealed by the Lissajous curves. At some characteristic $%
\gamma _{0}$, higher harmonics drop to almost zero indicating an unexpected
harmonic response even though the sample is under non-linear shear. Finally,
at a $\varphi $\ dependent intermediate $Pe_{\omega }$ regime the sample is
affected by both mechanisms (Brownian- and collision-induced yielding)
resulting in a double peak of $G_{1}^{\prime \prime }$\ and two minima of
higher harmonics, a feature that is detected for the first time in simple
hard sphere glasses.

The rich mechanical response of a model hard sphere glass revealed here in
oscillatory shear as a function of frequency, bridging the Brownian and
non-Brownian regimes, may provide insights for the understanding of systems
with more complicated interparticle interactions such as pastes, slurries,
particle gels, jammed emulsions and metallic glasses under a wide range of
conditions.

We thank A. B. Schofield for the particles and acknowledge funding from EU
FP7-Infrastructures 'ESMI' (CP\&CSA-2010-262348) and Greek "Thales" project
"COVISCO".

\bibliographystyle{apsrev}
\bibliography{Oscillatory_Nick_lett}

\bigskip

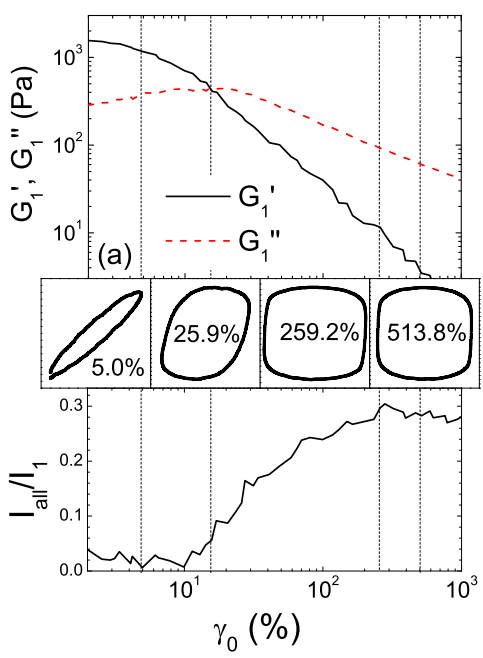
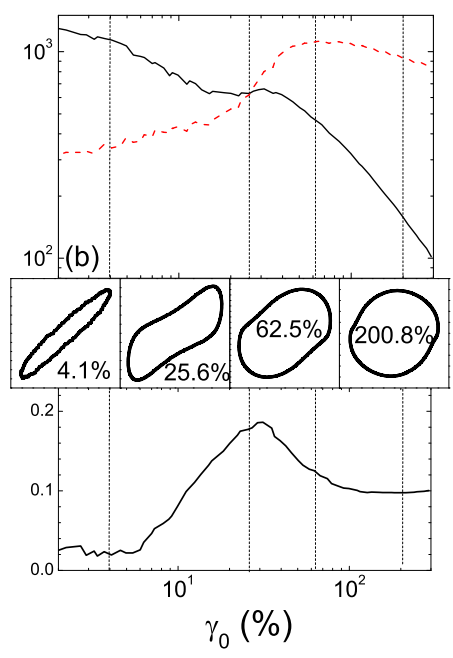
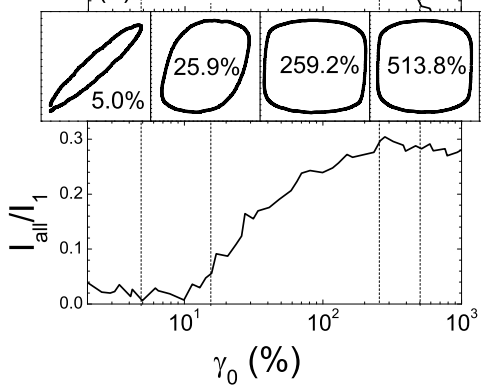
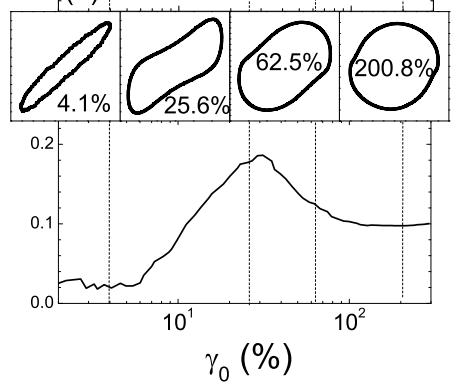
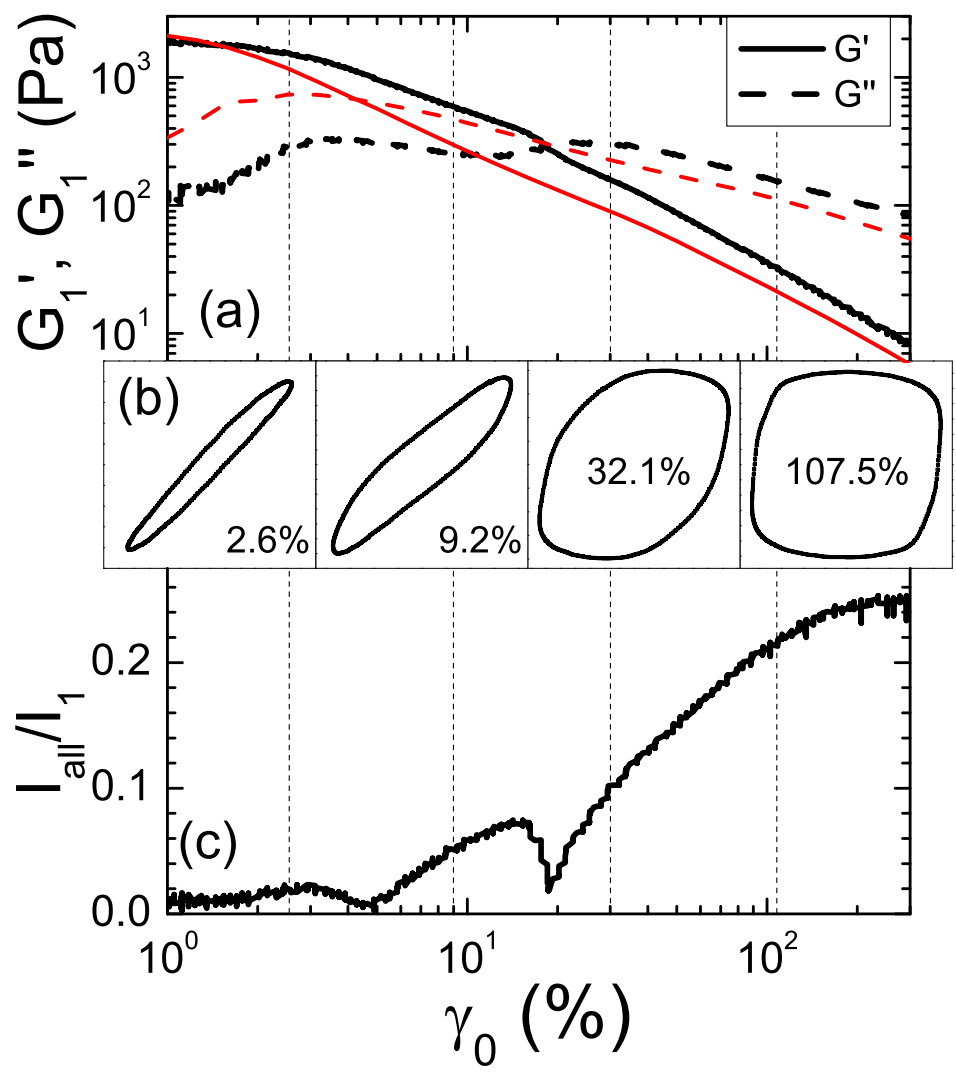
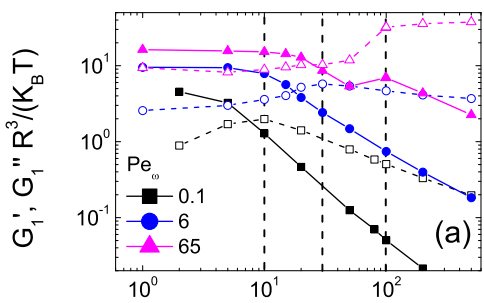
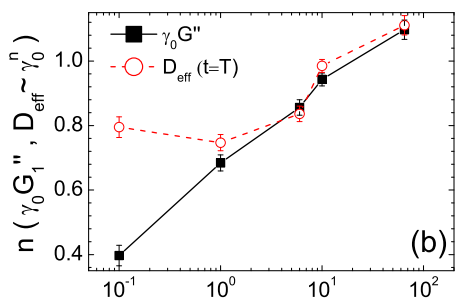
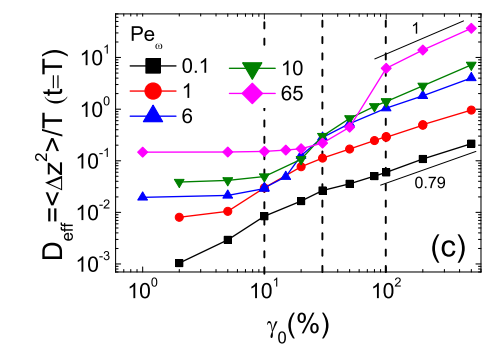
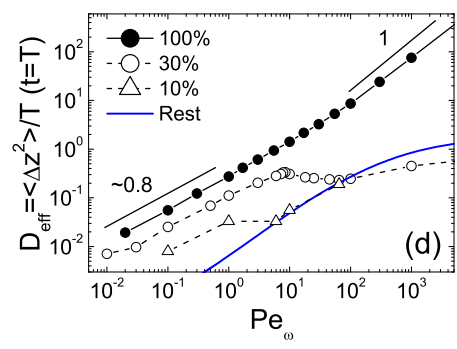
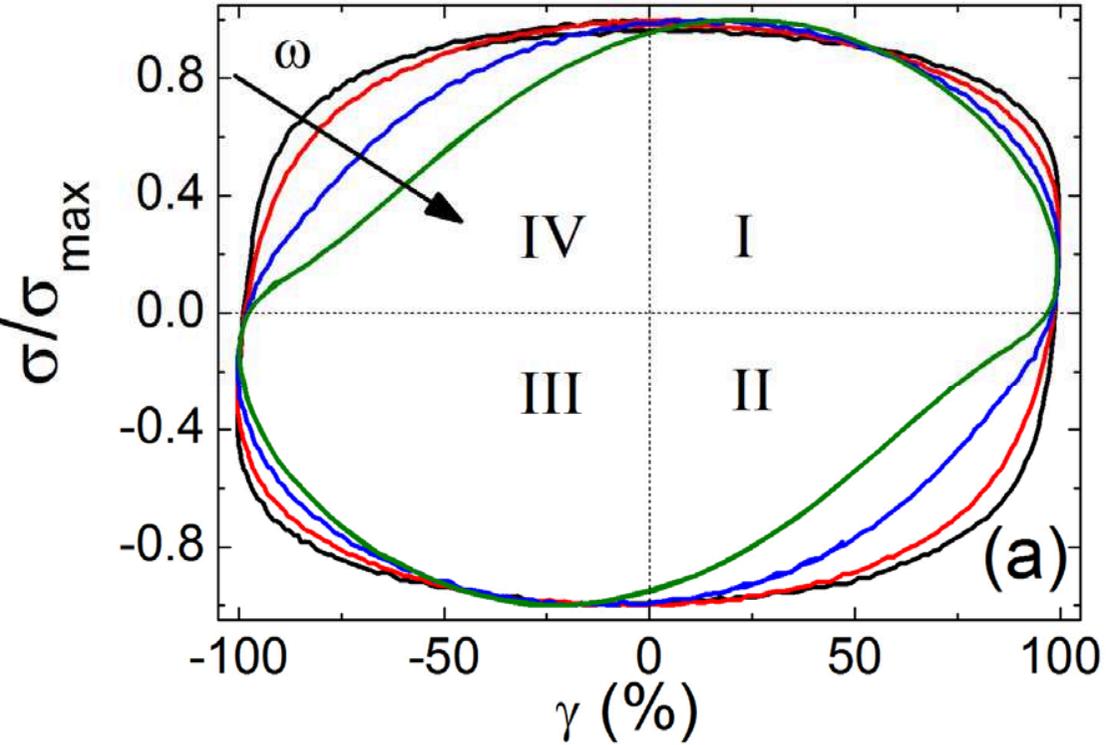
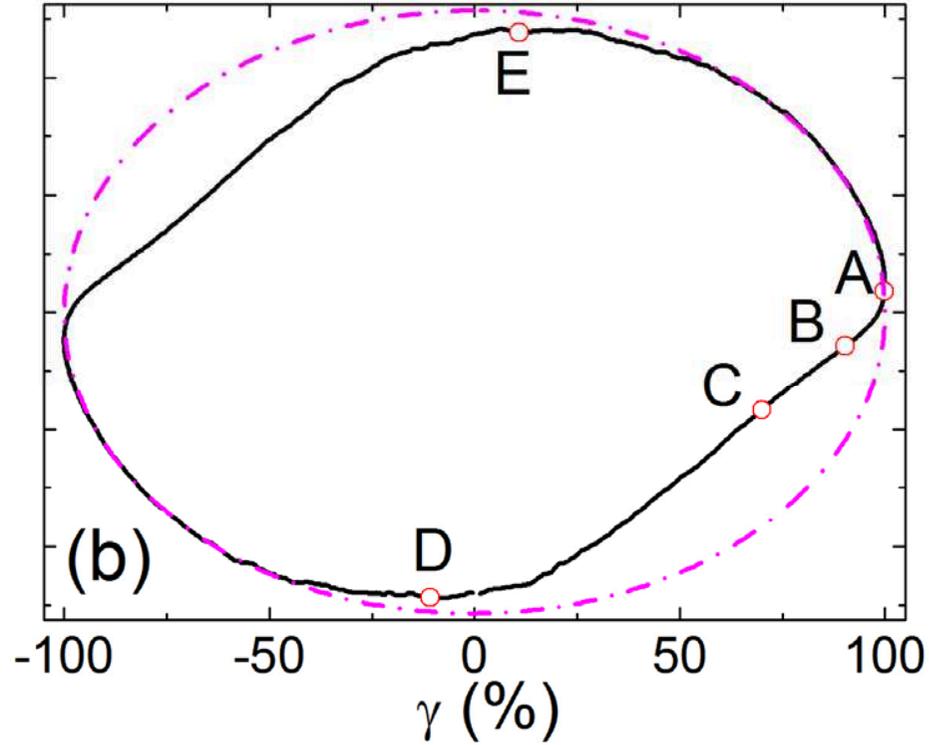
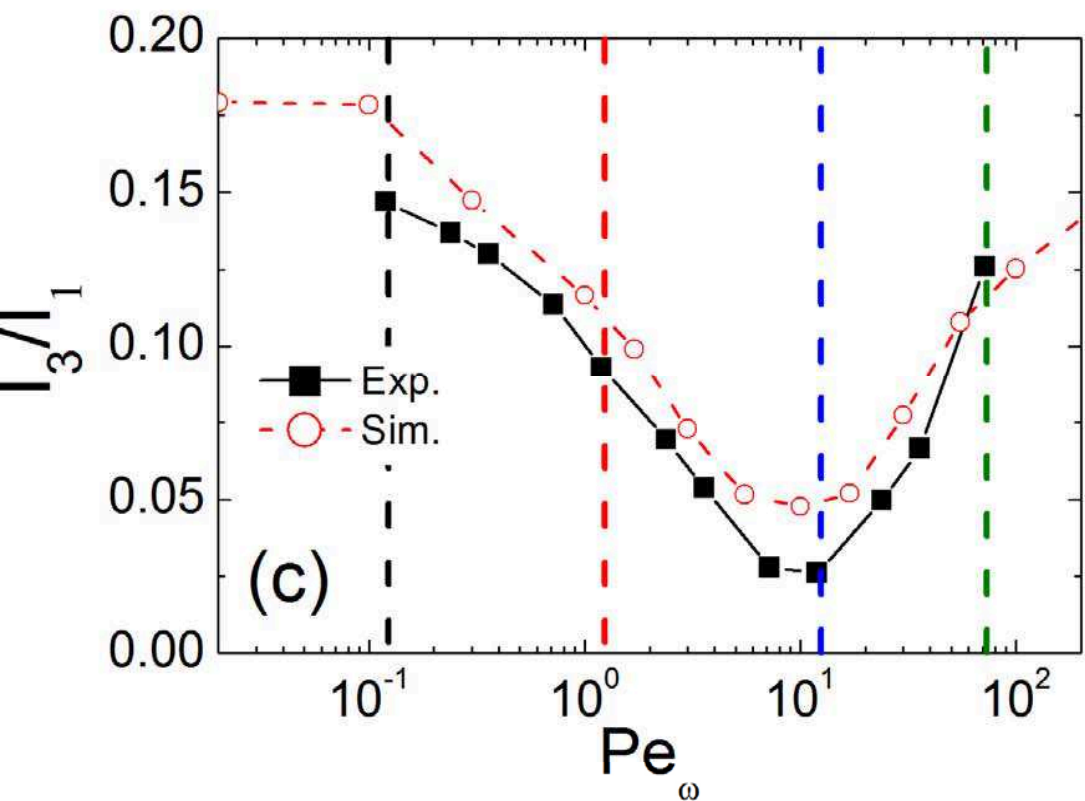
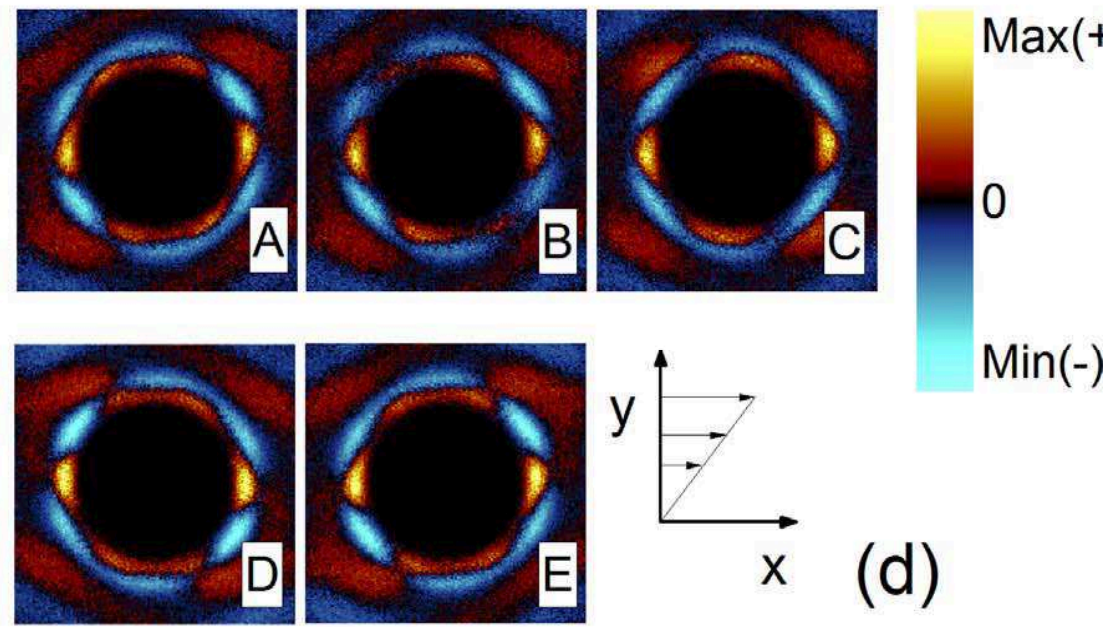
\end{document}